\documentclass[twocolumn,showpacs,preprintnumbers,amsmath,amssymb,nofootinbib]{revtex4}
\usepackage{amsthm}

\newtheorem{thm}{Theorem}

\renewcommand{\d}{{\,\rm d}}
\newcommand{\e}{{\,\rm e}}

\begin{document}
\title{Constraint preserving boundary conditions for the linearized BSSN formulation}
\author{Alexander M.~Alekseenko}%
 \email{alexander.alekseenko@csun.edu}
 \homepage{http://www.csun.edu/~ama5348}
 \affiliation{Department of Mathematics, California State University Northridge, Northridge,
              California 91330}%
\date{May 16, 2004, revised March 26, 2005}

\begin{abstract}
We derive two sets of explicit algebraic constraint preserving
boundary conditions for the linearized BSSN system. The approach
can be generalized to inhomogeneous differential and evolution
conditions, the examples of which are given. The proposed
conditions are justified by an energy estimate on the original BSSN variables.
\end{abstract}

\pacs{04.20.Ex, 04.25.Dm}
\maketitle

\section{Introduction}

The widely used treatment of Einstein's equations in numerical
relativity is to cast them to the form of a nonlinear hyperbolic
system with constraints (e.g., \cite{AnY,FR,KST,BS98,SN95}) and
solve by employing sophisticated discretization techniques. In the
course of solution, the constraint part is either monitored, or
explicitly imposed. It was observed, that the solution of the
evolution part with no constraints produces a violation which
grows rapidly breaking computations in a short time
\cite{KST,SKLPT02}. An attempt to control constraint violation, by
projecting the solution, or by incorporating constraint quantities
in the evolution equations, results in a longer life time of
calculations as a rule of thumb (e.g.,
\cite{HLOPSK04,LSKPST04,AABSS00,KLLPSST01}). It was found in
\cite{LSKPST04}, that exponentially growing constraint violating
solutions converge to unstable solutions of the dynamic equations,
which suggests that the constraint violation is closely related to
loss of stability in the system.

An exact solution to evolution equations in the entire space has a
property that it satisfies constraint equations automatically as
long as it satisfies them initially. However, in numerical
simulations, because of the roundoff and truncation errors, one
cannot hope for automatic constraint compliance. Instead, care
must be taken to ensure that the inserted perturbations are small,
and remain small during the evolution.

The behavior of the solution can be improved significantly
\cite{LSKPST04,KLS04} by introducing special sets of boundary
data, the so-called constraint-preserving boundary conditions, or
conditions that imply trivial evolution of constraints. Several
sets of such data were proposed for various first order
formulations of Einstein's equations (e.g.,
\cite{CPST03,AT03,KLS04,GMG04b,LSKPST04}).\footnote{An approach not
involving first order reduction has been proposed in \cite{FG03}
where boundary conditions were constructed by projecting
Einstein's equations on timelike boundaries.}
These conditions are typically written as a system of partial
differential equations restricted to the boundary, and in cases
when the equations are time dependent and decouple from the bulk
system, the equations may be integrated in time to produce regular
Dirichlet data that is compatible with constraints \cite{CPST03}.

In this work, two sets of well-posed homogeneous algebraic
constraint-preserving boundary conditions for the linearized
Baumgarte--Shapiro--Shibata--Nakamura formulation \cite{SN95,BS98}
are constructed. As is common, our derivation starts from
considering the evolution equations for constraint quantities and
looks for sets of data for the variables of the main system that
guarantee zero Dirichlet data for the constraint quantities. The
procedure is similar to the procedure found in \cite{CPST03} but
a) does not employ reduction to first order, and b) does not
involve integration of equations in time along the boundary.
Instead, following \cite{AT03,T04}, we rewrite the equations in a
special form to find well-posed constraint-preserving boundary
conditions by direct inspection. The approach can be generalized
to produce boundary conditions of the evolving type (see,
\cite{CPST03}) and the differential type \cite{LSKPST04,KLS04}. To
further justify the proposed conditions, we derive an energy
estimate for the nonlinear BSSN system with boundaries extending
the results of \cite{GMG04a,GMG04b}, and demonstrate that the nonlinear estimate
has the same boundary terms as in the linearized case.

In Section~2 we recall the derivation of the BSSN formulation and
use the opportunity to discuss choices of lapse and shift most
commonly found in numerical relativity. Section~3 describes the
linearization of the BSSN equations. In Section~4, the constraint
preserving boundary conditions are derived, and the generalization
to evolving and differential boundary conditions is discussed. In
Sections~5, and~6 the initial-boundary problem is defined using
the derived conditions. Also, in Section~6, a set of boundary
conditions for the dynamic part of the BSSN system, which is a
system first order in time, second order in space, is formulated.
Section~7 describes an energy estimate for the BSSN system not
involving first order in space or second order in time reductions.

\section{The trace-free decomposition of the ADM system}

To point out some facts about the nature and properties of the
BSSN formulation (see \cite{SN95,BS98} also, some special cases
are in \cite{ABDKPST03,AABSS00}), let us briefly recall the
derivation in the case of vacuum fields where the right hand side
of Einstein's equation is zero.\footnote{A reader not interested
in the BSSN derivation may proceed to Section~3 where the
linearized system is listed.}

The derivation starts from the
Arnowitt--Deser--Misner $3+1$ decomposition \cite{ADM,York},
\begin{gather}
\label{ADM1}
\partial_{0} h_{ij}
= -2 a k_{ij}+2h_{l(i}\partial_{j)}b^{l}, \\
\partial_{0} k_{ij}
= a[R_{ij} + (k_l^l) k_{ij}-2k_{il}k^{l}_{j}] \nonumber \\
\label{ADM2}
  \hspace{6mm}{}+k_{il}\partial_{j}b^{l}+k_{lj}\partial_{i}b^{l}-D_{i}D_{j}a, \\
\label{ADM3}
R_i^i+(k_i^i)^2 - k_{ij}k^{ij} =0,
\\
\label{ADM4}
D^j k_{ij}-D_i k_j^j = 0.
\end{gather}
Here $a$ denotes the lapse, the $b_i$ are the components of the
shift vector $b$, $h_{ij}$ are the components of the spatial metric $h$.
The components of the 4-dimensional metric $g$ are given by
\begin{equation*}
g_{00}=-a^2+b_ib_jh^{ij},\quad
g_{0i}=b_i,\quad
g_{ij}=h_{ij}.
\end{equation*}
$h^{ij}$ denotes the matrix inverse to $h_{ij}$, indices are
raised and traces taken with respect to the spatial metric;
$\partial_{0}:=(\partial_{t}-b^{s}\partial_{s})$ is the convective
derivative, $D_i$ is the covariant derivative operator associated
to the spatial metric; the extrinsic curvature $k_{ij}$ is defined
by equation (\ref{ADM1}); we assume that global Cartesian
coordinates $t=x_0$, $x_{1}$, $x_2$, $x_{3}$ are specified;
$R_{ij}$ are the components of the spatial Ricci tensor
\begin{eqnarray*}
R_{ij}
&=&\frac{1}{2}h^{pq}(\partial_{p}\partial_{j}h_{iq}+
   \partial_{i}\partial_{p}h_{qj}-\partial_{p}\partial_{q}h_{ij}-
   \partial_{i}\partial_{j}h_{pq}) \\
& &{}+h^{pq}h^{rs}(\Gamma_{ips}\Gamma_{qjr}-\Gamma_{pqs}\Gamma_{ijr}).
\end{eqnarray*}
where $\Gamma_{ijk}$ are the spatial Christoffel symbols defined by
$\Gamma_{ijk}=(\partial_{i}h_{kj}+\partial_{j}h_{ik}-\partial_{k}h_{ji})/2$.
%
%

The operator $R_{ij}$ in (\ref{ADM2}) contains second order
spatial derivatives of unknown fields and is very difficult to
analyze. As a result, it is difficult to judge about properties of
equation (\ref{ADM2}), and properties of $k_{ij}$ in general.
However, for the trace of the extrinsic curvature $k=k^{i}_{i}$,
the situation is different. Taking the trace of (\ref{ADM2}) and
using (\ref{ADM1}), (\ref{ADM3}) we find
\begin{equation}
\label{k1}
\partial_{0} k^{i}_{i} = ak^{li}k_{li} - D^{l}D_{l}a.
\end{equation}
The remarkable simplicity of the latter equation suggests to
separate the evolution of the trace of the extrinsic curvature
from the system. Specifically, we introduce the trace of the
extrinsic curvature $k=k^{i}_{i}$ and the trace-free part of the
extrinsic curvature $A_{ij}=k_{ij}-(1/3)h_{ij}k$ as new variables.
Then (\ref{k1}) yields
\begin{equation}
\label{k2}
\partial_{0} k = \frac{1}{3}ak^{2} + aA^{lm}A_{lm} - D^{l}D_{l}a.
\end{equation}
Unless the lapse function $a$ is chosen with care, equation
(\ref{k2}) is expected to be unstable. For example, for a
spatially independent lapse and zero shift vector, equation
(\ref{k2}) yields an estimate $\partial_{t} k \ge  \frac{1}{3}a
k^{2}$ which implies that $k \ge [(1/3)\int_{0}^{t}a(\tau)\d \tau
+1/k(0)]^{-1}$, or that the solution $k$ is unbounded in a finite
time, which is a well-known example of a coordinate singularity.
The problem can be solved, for example, by imposing maximal
slicing in the BSSN formulation \cite{ABDKPST03}
\begin{equation*}
D^{l}D_{l} a = a k^{lm}k_{lm}.
\end{equation*}
With this condition equation (\ref{k2}) reduces to $\partial_{0}k=0$.

Alternatively, it is often proposed to use harmonic slicing\footnote{Harmonic slicing is a
particular case of Bona-Mass\'{o} family of $k$-driving slicing
conditions $(\partial_{t} - b^{l}D_{l})a = -a^{2} f(a)k$, $f(a)>0$ \cite{AB01,BMSS95}}
\cite{BS98,ABDKPST03} which corresponds to setting
\begin{equation}
\label{a2}
\partial_{0}a = -a^2 k.
\end{equation}

The equation on $A$ is obtained from (\ref{ADM2}),
(\ref{k2}), and (\ref{ADM1}) as
\begin{eqnarray}
\label{A1}
\partial_{0}A_{ij}
&=& aR_{ij}+\frac{1}{3}akA_{ij}-2aA_{il}A^{l}_{j}
    +\frac{2}{9}ak^2h_{ij}\nonumber \\
& &{}-\frac{1}{3}aA^{lm}A_{lm}h_{ij}+A_{il}\partial_{j}b^{l}+A_{jl}\partial_{i}b^{l}
   \nonumber \\
& &{}- D_{i}D_{j}a + \frac{1}{3}h_{ij}D^{l}D_{l}a.
\end{eqnarray}
To proceed with the derivation we need a splitting for the spatial
metric $h$ compatible to the splitting of $k_{ij}$ into $k$ and
$A_{ij}$. In the BSSN formulation, the desired splitting is achieved
by introducing the conformal factor
$\varphi=(1/12)\ln(\det(h_{ij}))$ and the conformal metric
$\tilde{h}_{ij}=\e^{-4\varphi} h_{ij}$,
$\tilde{h}^{ij}=\e^{4\varphi} h^{ij}$. Using Leibnitz formula for
differentiating the determinant of a matrix
\begin{equation}
\label{det}
\partial \det(h_{ij})= \det(h_{ij})h^{lm}\partial h_{ml}
\end{equation}
one finds that the derivative of the conformal metric is trace-free:
\begin{equation}
\label{tildeh1}
\partial \tilde{h}_{ij}=\e^{-4\varphi}[\partial h_{ij}-\frac{1}{3}h_{ij}h^{lm}\partial h_{lm}].
\end{equation}
By applying operator $\partial_{0}$ on the definition of $\varphi$ and
using (\ref{det}), (\ref{ADM1}) we get the second equation of our
system
\begin{equation}
\label{phi1}
\partial_{0} \varphi = -\frac{1}{6} a k + \frac{1}{6} \partial_{l} b^{l}.
\end{equation}
Now using (\ref{tildeh1}) and (\ref{ADM1}) we obtain the third equation
\begin{equation}
\label{tildeh2}
\partial_0 \tilde{h}_{ij}
= -2a\tilde{A}_{ij}+2\tilde{h}_{l(i}\partial_{j)} \tilde{b}^{l} - \frac{2}{3}\tilde{h}_{ij}\partial_{l}\tilde{b}^{l},
\end{equation}
where $\tilde{A}_{ij}=\e^{-4\varphi}A_{ij}$, $\tilde{b}_{i}=
\e^{-4\varphi} b_{i}$ are the conformal analogs of the variables
$A$ and $b$. Beginning from the last equation, indices are lowered
and raised with the conformal metric $\tilde{h}_{ij}$ and its
inverse $\tilde{h}^{ij}=\e^{4\varphi}h^{ij}$ (in this case
$b^{s}=\tilde{b}^{s}$, and it is easy to redefine
$\partial_{0}=\partial_{t}-\tilde{b}^{s}\partial_{s}$).

The remaining two equations can be obtained from (\ref{A1}) which can be rewritten in terms of $\tilde{A}$ as
\begin{eqnarray}
\label{tildeA1}
\partial_{0}\tilde{A}_{ij}
&=& a\e^{-4\varphi}R_{ij}+a(k\tilde{A}_{ij}-2\tilde{A}_{il}\tilde{A}^{l}_{j}
    +\frac{2}{9}k^2\tilde{h}_{ij}\nonumber\\
& & -\frac{1}{3}\tilde{A}^{lm}\tilde{A}_{lm}\tilde{h}_{ij})+\tilde{A}_{il}\partial_{j}\tilde{b}^{l}
    +\tilde{A}_{jl}\partial_{i}\tilde{b}^{l}-\frac{2}{3} \tilde{A}_{ij} \partial_{l}\tilde{b}^{l} \nonumber \\
& &{}- \e^{-4\varphi}D_{i}D_{j}a  + \e^{-4\varphi}\frac{1}{3}\tilde{h}_{ij}D^{l}D_{l}a.
\end{eqnarray}
The Ricci tensor in terms of the conformal metric reads \cite{BS98,ABDKPST03}:
\begin{eqnarray}
\label{Ricci2}
R_{ij}
&=&\frac{1}{2}\tilde{h}^{pq}(\partial_{p}\partial_{j}\tilde{h}_{iq}+
   \partial_{i}\partial_{p}\tilde{h}_{qj}-\partial_{p}\partial_{q}\tilde{h}_{ij}-
   \partial_{i}\partial_{j}\tilde{h}_{pq}) \nonumber \\
& &{}-2\tilde{D}_{i}\tilde{D}_{j}\varphi
    -2\tilde{h}_{ij}\tilde{h}^{pq}\tilde{D}_{p}\tilde{D}_{q}\varphi
   \nonumber \\
& &{}+\tilde{h}^{pq}\tilde{h}^{rs}(\tilde{\Gamma}_{ips}\tilde{\Gamma}_{qjr}
   -\tilde{\Gamma}_{pqs}\tilde{\Gamma}_{ijr}) \nonumber \\
& &{}+4\partial_{i}\varphi\partial_{j}\varphi
   -4\tilde{h}_{ij}\tilde{h}^{pq}\partial_{p}\varphi\partial_{q}\varphi.
\end{eqnarray}
Here
$\tilde{\Gamma}_{ijk}=(\partial_{i}\tilde{h}_{kj}+\partial_{j}\tilde{h}_{ik}-\partial_{k}\tilde{h}_{ji})/2$;
$\tilde{D}_{i}v_{j}=\partial_{i}v_{j}-\tilde{h}^{pq}\tilde{\Gamma}_{ijp}v_{q}$ is the covariant derivative associated
with the conformal metric. The first line in (\ref{Ricci2}) can be
rewritten
\begin{eqnarray}
\label{Ricci3}
R_{ij}
&=& -\frac{1}{2}\tilde{h}^{pq}\partial_{p}\partial_{q}\tilde{h}_{ij}+\partial_{(i}\tilde{h}^{pq}\tilde{\Gamma}_{|pq|j)}
    + \tilde{\Gamma}_{pq(i}\partial_{j)}\tilde{h}^{pq}\nonumber\\
& &{}-2\tilde{D}_{i}\tilde{D}_{j}\varphi
   -2\tilde{h}_{ij}\tilde{h}^{pq}\tilde{D}_{p}\tilde{D}_{q}\varphi
   + {}    \ldots
\end{eqnarray}
This suggests to introduce a new variable
\begin{equation}
\label{tildeGamma1}
\tilde{\Gamma}_{j}= \tilde{h}^{pq} \tilde{\Gamma}_{pqj}=\tilde{h}^{pq}\partial_{p}\tilde{h}_{qj}.
\end{equation}
Substituting (\ref{Ricci3}) in (\ref{tildeA1}) one gets
the fourth evolution equation
\begin{eqnarray}
\label{tildeA2}
\partial_{0}\tilde{A}_{ij}
&=& -\frac{1}{2}a\e^{-4\varphi}\tilde{h}^{pq}\partial_{p}\partial_{q}\tilde{h}_{ij} +
    a\e^{-4\varphi}\partial_{(i}\tilde{\Gamma}_{j)}
    \nonumber \\
& & -2 a\e^{-4\varphi}\tilde{D}_{i}\tilde{D}_{j}\varphi
    -2 a\e^{-4\varphi}\tilde{h}_{ij}\tilde{h}^{pq}\tilde{D}_{p}\tilde{D}_{q}\varphi
    \nonumber \\
& &{} - \e^{-4\varphi}D_{i}D_{j}a  + \frac{1}{3}\e^{-4\varphi}\tilde{h}_{ij}D^{l}D_{l}a +
    W_{ij},\qquad
\end{eqnarray}
where
\begin{eqnarray*}
W_{ij}
&=& a \e^{-4\varphi}\tilde{\Gamma}_{pq(i}\partial_{j)}\tilde{h}^{pq}\nonumber\\
& &{}+a \e^{-4\varphi} \tilde{h}^{pq}\tilde{h}^{rs}(\tilde{\Gamma}_{ips}\tilde{\Gamma}_{qjr}
                                   -\tilde{\Gamma}_{pqs}\tilde{\Gamma}_{ijr}) \nonumber \\
& &{}+ 4 a \e^{-4\varphi} \partial_{i}\varphi\partial_{j}\varphi
    - 4 a \e^{-4\varphi}\tilde{h}_{ij}\tilde{h}^{pq}\partial_{p}\varphi\partial_{q}\varphi
    \nonumber \\
& &{}+a(k\tilde{A}_{ij}-2\tilde{A}_{il}\tilde{A}^{l}_{j}
     +\frac{2}{9}k^2\tilde{h}_{ij}-\frac{1}{3}\tilde{A}^{lm}\tilde{A}_{lm}\tilde{h}_{ij}) \nonumber \\
& & {}+\tilde{A}_{il}\partial_{j}\tilde{b}^{l}+\tilde{A}_{jl}\partial_{i}\tilde{b}^{l}
      -\frac{2}{3} \tilde{A}_{ij} \partial_{l}\tilde{b}^{l}.
\end{eqnarray*}

The evolution equation on $\tilde{\Gamma}$ is obtained by
differentiating its definition and using the momentum constraint.
Namely, we apply operator $\partial_{0}$ on (\ref{tildeGamma1}) to
get
\begin{eqnarray}
\label{tildeGamma2}
\partial_{0}\tilde{\Gamma}_{i}&=&
 -2a\partial^{l}\tilde{A}_{li}
 -2\tilde{h}^{pq}(\partial_{p}a)\tilde{A}_{qi}
 +2a\tilde{A}^{pq}(\partial_{p}\tilde{h}_{qi}) \nonumber \\
& &{} +\Gamma_{l}\partial_{i}\tilde{b}^{l}
 +\frac{1}{3}\partial_{i}\partial_{l}\tilde{b}^{l}
 +\tilde{h}_{li}\partial^{s}\partial_{s}\tilde{b}^{l}
\end{eqnarray}
Then we notice that
$h^{pq}D_{p}A_{iq}=\partial^{p}\tilde{A}_{pi}-\tilde{\Gamma}_{s}\tilde{A}^{s}_{i}
+6(\partial_{s}\varphi)\tilde{A}^{s}_{i}$, and thus (\ref{ADM4}) reduces to
\begin{equation*}
\partial^{p}\tilde{A}_{pi}-\frac{2}{3}\partial_{i} k -\tilde{\Gamma}_{s}\tilde{A}^{s}_{i}
+6(\partial_{s}\varphi)\tilde{A}^{s}_{i} = 0.
\end{equation*}
Solving this equation for $\partial^{l}\tilde{A}_{l}$ and
substituting the result in (\ref{tildeGamma2}) we derive the fifth equation
of the BSSN system
\begin{equation}
\label{tildeGamma3}
\partial_{0}\tilde{\Gamma}_{i}=
 -\frac{4}{3}a\partial_{i} k + S_{i},
\end{equation}
where
\begin{eqnarray*}
S_{i}
 &=&-2a\tilde{\Gamma}_{s}\tilde{A}^{s}_{i}
    +12a(\partial_{s}\varphi)\tilde{A}^{s}_{i}
    -2\tilde{h}^{pq}(\partial_{p}a)\tilde{A}_{qj}\\
 & &{}+2a\tilde{A}^{pq}(\partial_{p}\tilde{h}_{qj})
 +\Gamma_{l}\partial_{j}\tilde{b}^{l}
 +\frac{1}{3}\partial_{i}\partial_{l}\tilde{b}^{l}
 +\tilde{h}_{lj}\partial^{s}\partial_{s}\tilde{b}^{l}.
\end{eqnarray*}
Equations (\ref{k2}), (\ref{phi1}), (\ref{tildeh2}),
(\ref{tildeA2}), and (\ref{tildeGamma3}) constitute the core of
the BSSN formulation. These equations are usually supplemented by
one or more equations describing the choice of the gauge
functions. Thus, in most cases the lapse and the shift are not
known but dynamically depend on the metric and other quantities.
In this work, we will assume the harmonic lapse condition
(\ref{a2}). Further we consider either a prescribed shift $b_{i}$
or a shift that follows from the gamma-freezing condition
$\partial_{t}\tilde{\Gamma}_{i}=0$ (see, for example, \cite{ABDKPST03}).

\section{Linearization around Minkowski space}

Minkowski spacetime in Cartesian coordinates is represented by the
trivial solution to ADM system: $h_{ij}=\delta_{ij}$, $k_{ij}=0$,
$a=1$, $b_i=0$. Consider perturbations of ADM variables around the
Minkowski spacetime: $h_{ij}=\delta_{ij}+\gamma_{ij}$,
$k_{ij}=\kappa_{ij}$, $a=1+\alpha$, $b_i=\beta_i$, with the
$\gamma_{ij}$, $\kappa_{ij}$, $\alpha$, and $\beta_i$ supposed to
be small. Substituting these expressions into the definitions of
the BSSN variables and neglecting terms of second and higher order
in perturbations we get
\begin{gather}
\det(h_{ij})=1+\gamma^{l}_{l},\quad
\varphi=\frac{1}{12}\gamma^{l}_{l}, \nonumber \\
\e^{-4\varphi}=1-\frac{1}{3}\gamma^{l}_{l},\quad
\e^{4\varphi}=1+\frac{1}{3}\gamma^{l}_{l}, \nonumber \\
\tilde{h}_{ij}=\delta_{ij}+\gamma_{ij}-\frac{1}{3}\delta_{ij}\gamma^{l}_{l}
=:\delta_{ij}+\tilde{\gamma}_{ij}, \nonumber\\
k=\kappa=:\kappa^{l}_{l},\quad
A_{ij}=\tilde{A}_{ij}=\kappa_{ij}-\frac{1}{3}\delta_{ij}\kappa, \nonumber \\
\label{linvars1}
\Gamma_{i}=:\tilde{\Gamma}_{i}=\partial^{l}\tilde{\gamma}_{il}.
\end{gather}
Substituting these quantities in equations (\ref{phi1}), (\ref{tildeh2}), (\ref{k2}), (\ref{tildeA2}),
(\ref{tildeGamma3}), and (\ref{a2}) and ignoring the terms which are at least
second order in $\varphi$, $\tilde{\gamma}_{ij}$, $\kappa$, $A_{ij}$,
and $\Gamma_{i}$ we derive the linearized BSSN system
\begin{eqnarray}
\label{LBSSN1}
\partial_{t}\varphi
 &=& -\frac{1}{6}\kappa+\frac{1}{6}\partial^{l}\beta_{l}, \\
\label{LBSSN2}
\partial_{t}\alpha
 &=& -\kappa, \\
\label{LBSSN3}
\partial_{t}\kappa
 &=&-\partial^{l}\partial_{l}\alpha, \\
\label{LBSSN4}
\partial_{t} \tilde{\gamma}_{ij}
 &=& -2A_{ij}+2\partial_{(i}\beta_{j)}
     -\frac{2}{3}\delta_{ij} \partial^{l} \beta_{l},\\
\label{LBSSN5}
\partial_{t}A_{ij}
 &=& -\frac{1}{2}\partial^{l}\partial_{l} \tilde{\gamma}_{ij}
     + \partial_{(i}\Gamma_{j)} - 2\partial_{i}\partial_{j}\varphi \nonumber \\
 & &{}-2\delta_{ij}\partial^{l}\partial_{l}\varphi
      -\partial_{i}\partial_{j}\alpha+\frac{1}{3}\delta_{ij}\partial^{l}\partial_{l}\alpha, \\
\label{LBSSN6}
\partial_{t}\Gamma_{i}
 &=& -\frac{4}{3}\partial_{i}\kappa + \frac{1}{3}\partial_{i}\partial^{p}\beta_{p}
     +\partial^{p}\partial_{p}\beta_{i}.
\end{eqnarray}
Notice that the linearized harmonic lapse condition is included in
this system in the form of equation (\ref{LBSSN2}). In fact, this
condition will be used in the hyperbolic reduction which does not seem possible in general.

Linearization of Hamiltonian and the momentum constraint equations yields
correspondingly
\begin{gather}
\label{LBSSN7}
\partial^{l}\partial^{j}\tilde{\gamma}_{lj}-8\partial^{l}\partial_{l}\varphi=0,\quad
\partial^{l}\Gamma_{l}-8\partial^{l}\partial_{l}\varphi=0, \\
\label{LBSSN8}
\partial^{l}A_{il}-\frac{2}{3}\partial_{i}\kappa=0.
\end{gather}
Hamiltonian constraint appears in two versions since it can be
written both in terms of $\tilde{\gamma}$ and $\Gamma$. Also,
introducing the new variable $\Gamma$ entails an artificial
constraint
\begin{equation}
\label{LBSSN9}
\Gamma_{i}=\partial^{l}\tilde{\gamma}_{il}.
\end{equation}

The linearized problem then consists of determining $\varphi$,
$\alpha$, $\kappa$, $\tilde{\gamma}$, $A$, $\Gamma$ from equations
(\ref{LBSSN1})--(\ref{LBSSN6}) provided initial data and
admissible boundary data. The constraint equations
(\ref{LBSSN7})--(\ref{LBSSN8}) may or may not be imposed during
the evolution. The initial data $\varphi(0)$, $\kappa(0)$,
$\tilde{\gamma}(0)$,  $A(0)$, $\Gamma(0)$ is determined from
$\gamma(0)$ and $\kappa(0)$ using (\ref{linvars1}). It can be
checked that if $\gamma(0)$ and $\kappa(0)$ satisfy the linearized
Hamiltonian and momentum constraints in the ADM
system,\footnote{The linearized Hamiltonian and momentum
constraints in the ADM system are
$\partial^{l}\partial^{i}\gamma_{li}-\partial^{l}\partial_{l}\gamma^{i}_{i}=0$
and $\partial^{l}\kappa_{il}-\partial_{i}\kappa^{l}_{l}=0$.}
then $\varphi(0)$, $\kappa(0)$,
$\tilde{\gamma}(0)$,  $A(0)$, $\Gamma(0)$ satisfy the constraint
equations (\ref{LBSSN7})--(\ref{LBSSN9}).

\section{Constraint preserving boundary conditions}

The BSSN system is a constrained evolution system in the sense
that it has the dynamic part (\ref{LBSSN1})--(\ref{LBSSN6}) and
the constraint part (\ref{LBSSN7})--(\ref{LBSSN9}). It was assumed
for a long time, that for the right boundary data, a solution to
(\ref{LBSSN1})--(\ref{LBSSN6}) will satisfy the constraints
automatically once it satisfies them initially, however, examples
of such data were constructed only recently \cite{GMG04b} for a
first order reduction. Still it remains a question, which we are
trying to address in this paper, whether a set of constraint
preserving boundary conditions can be proposed for the original
BSSN variables.

Let us notice that, for a solution of
(\ref{LBSSN1})--(\ref{LBSSN6}), constraints (\ref{LBSSN7}) and
(\ref{LBSSN9}) are consequences of (\ref{LBSSN8}), so we can focus
just on the last one. Indeed, in view of (\ref{LBSSN1}),
(\ref{LBSSN5}), and (\ref{LBSSN6}), the following equations hold
for the time derivative of (\ref{LBSSN7})
\begin{gather*}
\partial_{t}(\partial^{l}\partial^{j}\tilde{\gamma}_{lj}
-8\partial^{l}\partial_{l}\varphi) = \partial^{i}(\partial^{l}A_{il}
-\frac{2}{3}\partial_{i}\kappa), \\
\partial_{t}(\partial^{l}\Gamma_{l}-8\partial^{l}\partial_{l}\varphi)=0.
\end{gather*}
These equations state that both parts of (\ref{LBSSN7}) are
satisfied as long as they are satisfied initially, and
(\ref{LBSSN8}) is true. Similarly, if (\ref{LBSSN8}) is satisfied,
then the time derivative of (\ref{LBSSN9}) is zero in view of
(\ref{LBSSN4}), (\ref{LBSSN6}). Thus (\ref{LBSSN9}) remains zero
provided it is zero initially.

We will now construct boundary conditions for system
(\ref{LBSSN1})--(\ref{LBSSN6}) that preserve (\ref{LBSSN8}). We
introduce a new variable
\begin{equation}
\label{m1}
M_{i}=\partial^{l}A_{il}-\frac{2}{3}\partial_{i}\kappa.
\end{equation}
Equation (\ref{LBSSN8}) is satisfied iff $M_{i}=0$, in other
words, the condition in question must guarantee $M_{i}=0$.

By differentiating (\ref{m1}) twice in time and substituting time
derivatives of equations (\ref{LBSSN5}) and (\ref{LBSSN3}) for
$\partial^{2}_{t}A$, $\partial^{2}_{t}\kappa$, we derive (terms in
$\varphi$, $\alpha$, $\kappa$, $\beta$, $\Gamma$ cancel in view of
(\ref{LBSSN1})--(\ref{LBSSN6}))
\begin{equation}
\label{m2}
\partial^{2}_{t}M_{i}=\partial^{l}\partial_{l}M_{i}.
\end{equation}

Initial values $M(0)$ can be determined from
$A(0)$, $\kappa(0)$ using (\ref{m1}) and must be zero for physical initial
data. The initial values for $\partial_{t} M_{i}$ can be
calculated by differentiating (\ref{m1}) in time and substituting
(\ref{LBSSN3}) and (\ref{LBSSN5}) for $\partial_{t}\kappa$ and
$\partial_{t}A_{ij}$,
\begin{equation*}
\partial_{t}M_{i}
= -\frac{1}{2} \partial^{l}\partial_{l}\partial^{m}\tilde{\gamma}_{im}
    +\frac{1}{2}\partial_{i}\partial^{l}\Gamma_{l}
+\frac{1}{2}\partial^{l}\partial_{l}\Gamma_{i}
-4\partial_{i}\partial^{l}\partial_{l}\varphi.
\end{equation*}
It can be verified by substitution, that if $\tilde{\gamma}(0)$,
$\Gamma(0)$, $\varphi(0)$ satisfy (\ref{LBSSN7}),
(\ref{LBSSN9}), then $\partial_{t} M(0)_{i}=0$.

It remains to select the boundary conditions on $M$ that imply
trivial evolution of (\ref{m2}). However, we notice that the
boundary data on $M$ is expected not to be given freely but
determined by the boundary conditions on $A$ and $\kappa$, similar
to the way $M(0)$, $\partial_{t} M(0)$ is determined by the main
variables $A(0)$, $\kappa(0)$, $\tilde{\gamma}(0)$, $\Gamma(0)$,
and $\varphi(0)$. But we do not know how to specify the
boundary conditions on $A$ and $\kappa$ either! Here is the key:
we will select that data now by observing its relationship with
the boundary data on $M$ \cite{AT03,T04}. For the boundary
conditions, again, we expect both definition (\ref{m1}) and the
evolution equations (\ref{LBSSN1})--(\ref{LBSSN6}) to contribute
into the relationship.

We introduce scalar products
$(v_{i},u_{i})=\int_{\Omega}v_{i}u^{i} \d x \d y \d z$ and
$(\rho_{ij},\sigma_{ij})=\int_{\Omega}\rho_{ij}\sigma^{ij} \d x \d
y \d z$ for the spaces of vectorfields and matrixfields on $\Omega$
correspondingly. The $L_{2}$ norms naturally associated with these
scalar products, $\|u\|^{2}=(u_{i},u_{i})$ and
$\|\rho\|^{2}=(\rho_{ij},\rho_{ij})$, are denoted by $\| \cdot
\|$. We introduce the energy of system (\ref{m2}),
\begin{equation*}
\epsilon=\|\partial_{t} M\|^2+\|\partial_{l} M \|^2
\end{equation*}
If we prove that the energy $\epsilon$ remains zero at all times,
then, in view of the trivial initial data, this will imply that
$\|M\|\equiv 0$ (e.g, \cite{LU68}).

Differentiating $\epsilon$ in time, using the Green's First
Identity component-wise to transfer the spatial derivative in the
second term, and also (\ref{m2}), we obtain
\begin{equation}
\label{en0}
\partial_{t}\epsilon =\int_{\partial\Omega} (\frac{\partial}{\partial n} M_{i})(\partial_{t} M^{i}) \d
\sigma.
\end{equation}
The energy $\epsilon$ is not increasing if
\begin{equation}
\label{en1}
(\frac{\partial}{\partial n} M_{i})(\partial_{t} M^{i}) \le  0
\quad \mbox {on} \quad \partial\Omega.
\end{equation}
The desired boundary conditions on $A$ and $\kappa$ will follow
immediately once we rewrite (\ref{en1}) in terms of the main variables.

We assume that the boundary $\partial\Omega$ is a combination of
arbitrarily oriented planes and consider any of its faces. Let
vector $n_{i}$ be the unit vector perpendicular to the face,
$m_{i}$, $l_{i}$ complement $n_{i}$ to an orthonormal triple (for
example, $m_{i}$ is any unit vector parallel to the boundary and
$l_{i}$ is the cross product of $n_{i}$ and $m_{i}$,
$l_{i}=\varepsilon_{i}^{jk}n_{j}m_{k}=(n\times m)_{i}$). At the
flat boundary, the divergence of a vector field can be expressed
in terms of the directional derivatives along vectors $n$, $m$, and
$l$, as
\begin{equation}
\label{op1}
\partial^{i}v_{i} = n^{i} \frac{\partial}{\partial n} v_{i}+
m^{i}\frac{\partial}{\partial m} v_{i}+
l^{i}\frac{\partial}{\partial l} v_{i}.
\end{equation}
Similarly, the gradient of a scalar field $\psi$ reads
\begin{equation}
\label{op2}
\partial_{i}\psi = n_{i}\frac{\partial}{\partial n}\psi +
m_{i}\frac{\partial}{\partial m}\psi +
l_{i}\frac{\partial}{\partial l}\psi .
\end{equation}
Next we note that, at any point of the boundary a symmetric trace free
matrix is spanned by
\begin{equation*}
n_{(i}m_{j)},\
n_{(i}l_{j)},\
l_{(i}m_{j)},\
l_{i}l_{j}-m_{i}m_{j},\
2n_{i}n_{j}-l_{i}l_{j}-m_{i}m_{j}.
\end{equation*}
Introducing scalar functions
\begin{gather}
A1=2A^{ij}n_{(i}m_{j)},\quad
A2=2A^{ij}n_{(i}l_{j)},\nonumber \\
A3=2A^{ij}l_{(i}m_{j)},\quad
A4=\frac{1}{2} A^{ij}(l_{i}l_{j}-m_{i}m_{j}),\nonumber\\
\label{As1}
A5=\frac{1}{6} A^{ij}(2n_{i}n_{j}-l_{i}l_{j}-m_{i}m_{j})
\end{gather}
we restate $A$ as
\begin{eqnarray}
\label{A4}
A_{ij}
&=& A1\, (n_{(i}m_{j)})+ A2\, (n_{(i}l_{j)}) + A3\, (l_{(i}m_{j)}) \nonumber \\
& &{}+A4\, (l_{i}l_{j}-m_{i}m_{j}) +
A5\,(2n_{i}n_{j}-l_{i}l_{j}-m_{i}m_{j}).\nonumber\\
& &
\end{eqnarray}
Substituting (\ref{A4}) into (\ref{m1}) and using (\ref{op1}) and (\ref{op2}) to replace
partial derivatives with the directional derivatives, we get
\begin{align}
M_{i}
&= [\frac{1}{2}\frac{\partial}{\partial m}A1
             +\frac{1}{2}\frac{\partial}{\partial l}A2
             +2\frac{\partial}{\partial n}A5-\frac{2}{3}\frac{\partial}{\partial n} \kappa
             ] n_{i}
             \nonumber \\[1mm]
&{}+[\frac{1}{2}\frac{\partial}{\partial n} A1
             +\frac{1}{2}\frac{\partial}{\partial l}A3
             -\frac{\partial}{\partial m}A4
             -\frac{\partial}{\partial m}A5-\frac{2}{3}\frac{\partial}{\partial m} \kappa
             ] m_{i}
             \nonumber \\[1mm]
\label{m3}
&{}+[\frac{1}{2}\frac{\partial}{\partial n} A2
             +\frac{1}{2}\frac{\partial}{\partial m} A3
             +\frac{\partial}{\partial l}A4
             -\frac{\partial}{\partial l}A5-\frac{2}{3}\frac{\partial}{\partial l} \kappa
             ] l_{i}.
\end{align}
The last equation implies that
\begin{eqnarray}
\lefteqn{(\frac{\partial}{\partial n} M_{i})(\partial_{t} M^{i})}
& & \nonumber\\
&=& \frac{\partial}{\partial n}[\frac{1}{2}\frac{\partial}{\partial m}A1
             +\frac{1}{2}\frac{\partial}{\partial l}A2
             +2\frac{\partial}{\partial n}A5-\frac{2}{3}\frac{\partial}{\partial n} \kappa ]
             \nonumber \\[1mm]
& &{} \times \partial_{t} [\frac{1}{2}\frac{\partial}{\partial m}A1
             +\frac{1}{2}\frac{\partial}{\partial l}A2
             +2\frac{\partial}{\partial n}A5
             -\frac{2}{3}\frac{\partial}{\partial n}\kappa ]
             \nonumber \\[1mm]
&+&\frac{\partial}{\partial n}[\frac{1}{2}\frac{\partial}{\partial n} A1
             +\frac{1}{2}\frac{\partial}{\partial l}A3
             -\frac{\partial}{\partial m}A4
             -\frac{\partial}{\partial m}A5-\frac{2}{3}\frac{\partial}{\partial m} \kappa ]
             \nonumber \\[1mm]
& &{}\times \partial_{t} [\frac{1}{2}\frac{\partial}{\partial n}A1
             +\frac{1}{2}\frac{\partial}{\partial l}A3
             -\frac{\partial}{\partial m}A4
             -\frac{\partial}{\partial m}A5
             -\frac{2}{3}\frac{\partial}{\partial m}\kappa ]
             \nonumber \\[1mm]
&+&\frac{\partial}{\partial n}[\frac{1}{2}\frac{\partial}{\partial n}A2
             +\frac{1}{2}\frac{\partial}{\partial m}A3
             +\frac{\partial}{\partial l}A4
             -\frac{\partial}{\partial l}A5-\frac{2}{3}\frac{\partial}{\partial l} \kappa ]
             \nonumber \\[4mm]
& &{}\times \partial_{t} [\frac{1}{2}\frac{\partial}{\partial n}A2
             +\frac{1}{2}\frac{\partial}{\partial m}A3
             +\frac{\partial}{\partial l}A4
             -\frac{\partial}{\partial l}A5
             -\frac{2}{3}\frac{\partial}{\partial
             l}\kappa].\nonumber\\
& &
\label{m4}
\end{eqnarray}
Either of the two sets of boundary conditions imply $((\partial/\partial n)
M^{i})(\partial_{t} M_{i})=0$ on $\partial \Omega$:
\begin{gather}
A1=0,\quad A2=0,\quad
\frac{\partial}{\partial n}A3=0,\quad
\frac{\partial}{\partial n}A4=0,\nonumber \\
\label{ccbc1}
\frac{\partial}{\partial n}A5=0,\quad
\frac{\partial}{\partial n}\kappa=0, \\
\frac{\partial}{\partial n}A1=0,\quad
\frac{\partial}{\partial n}A2=0,\quad
A3=0,\quad A4=0, \nonumber \\
\label{ccbc2}
A5=0,\quad \kappa=0.
\end{gather}
In particular, (\ref{ccbc1}) eliminates the second multiplier
in the first term of (\ref{m4}) and the first multipliers in
the second and third terms (by commuting partial derivatives and
using (\ref{eqA1})). Condition (\ref{ccbc2}) is verified in a
similar way.

More examples of constraint-preserving boundary conditions can be
proposed by inspection of (\ref{m4}). For example, the condition
$M_{i}|_{\partial \Omega}=0$, according to (\ref{m3}),  is
equivalent to the set of differential boundary conditions that can
be implemented numerically \cite{LSKPST04,KLS04}:
\begin{align}
\frac{1}{2}\frac{\partial}{\partial m}A1
             +\frac{1}{2}\frac{\partial}{\partial l}A2
             +2\frac{\partial}{\partial n}A5-\frac{2}{3}\frac{\partial}{\partial n} \kappa
             &=0, \nonumber \\[1mm]
\frac{1}{2}\frac{\partial}{\partial n}A1
             +\frac{1}{2}\frac{\partial}{\partial l}A3
             -\frac{\partial}{\partial m}A4
             -\frac{\partial}{\partial m}A5-\frac{2}{3}\frac{\partial}{\partial m}
             \kappa &=0, \nonumber \\[1mm]
\label{ccbc3}
\frac{1}{2}\frac{\partial}{\partial n} A2
             +\frac{1}{2}\frac{\partial}{\partial m} A3
             +\frac{\partial}{\partial l}A4
             -\frac{\partial}{\partial l}A5-\frac{2}{3}\frac{\partial}{\partial l} \kappa
             &=0.
\end{align}
Namely, one could prescribe Dirichlet data on $A3$, $A4$, and
$\kappa$. Then, (\ref{ccbc3}) gives mixed conditions on $A1$,
$A2$, $A5$. The problem with this condition is that it is not
obvious if it leads to a well-posed evolution of (\ref{eqA1}) (the
next example, however, contains an idea on how the well-posedness
can be established).

Furthermore, one could have considered a combination of Neumann and
Dirichlet conditions
\begin{equation}
\label{ccbc4}
\frac{\partial}{\partial n} M_{i}n^{i}=0,\quad
M_{i}l^{i}=0,\quad M_{i}m^{i}=0.
\end{equation}
Applying $\partial/\partial m$ to the second,
$\partial/\partial l$ to the third equation of (\ref{ccbc3}) and
subtracting the results from the normal derivative of the first one, we derive
(using (\ref{eqA1}), (\ref{eqkappa1}) to eliminate
$(\partial^2/\partial n^2)$ derivatives) an evolution equation
defined on the boundary
\begin{gather}
2\frac{\partial^2}{\partial t^2}A5
- (\frac{\partial^2}{\partial l^2}+\frac{\partial^2}{\partial m^2})A5
- \frac{2}{3} \frac{\partial^2}{\partial t^2} \kappa
+ \frac{4}{3}(\frac{\partial^2}{\partial l^2}+\frac{\partial^2}{\partial
m^2})\kappa \nonumber \\
=
(\frac{\partial^2}{\partial l^2}-\frac{\partial^2}{\partial m^2})A4
+ \frac{\partial}{\partial l}\frac{\partial}{\partial m}A3.
\label{ccbc5}
\end{gather}
Equation (\ref{ccbc5}) can be used to find Dirichlet data on $A5$
and $\kappa$, provided Dirichlet data for $A3$ and $A4$ is
given (see \cite{CPST03}). Once $A5$ and $\kappa$ are known, the values of
$(\partial /\partial n)A1$, $(\partial /\partial n)A2$ can be
determined from the last two equations of (\ref{ccbc3}). The
corresponding inhomogeneous algebraic conditions on $A_{ij}$ then
would read
\begin{gather}
2(\partial /\partial n)A^{ij}n_{(i}m_{j)}=(\partial /\partial
n)A1,\nonumber\\
2(\partial /\partial n)A^{ij}n_{(i}l_{j)}=(\partial /\partial
n)A2,\nonumber \\
2A^{ij}l_{(i}m_{j)}=A3,\nonumber\\
(1/2)A^{ij}(l_{i}l_{j}-m_{i}m_{j})=A4,\nonumber\\
\label{ccbc6}
(1/6)A^{ij}(2n_{i}n_{j}-l_{i}l_{j}-m_{i}m_{j})=A5.
\end{gather}
The boundary conditions (\ref{ccbc6}) are constraint-preserving
and represent an analog of conditions introduced in \cite{CPST03,
GMG04b}.

\section{Evolution of $A$ and $\kappa$. Second order in time reduction}

We will argue now that the boundary conditions (\ref{ccbc1}) and
(\ref{ccbc2}) lead to a well-posed problem for the linearized BSSN
system. By differentiating equation (\ref{LBSSN5}) in time and
substituting (\ref{LBSSN4}) for $\partial_{t}\tilde{\gamma}_{ij}$
in the result, we obtain (terms in $\varphi$, $\alpha$, $\kappa$,
$\beta$, $\Gamma$ cancel in view of
(\ref{LBSSN1})--(\ref{LBSSN3}), (\ref{LBSSN6}))
\begin{equation}
\label{eqA1}
\partial^{2}_{t}A_{ij}=\partial^{l}\partial_{l}A_{ij}.
\end{equation}
We assume that the initial values $A(0)$ and $\partial_{t}A(0)$
are determined from (\ref{linvars1}) and (\ref{LBSSN5})
correspondingly, and that either of the conditions (\ref{ccbc1})
or (\ref{ccbc2}) is given at the domain boundary.

We introduce scalar products $(\mu,\nu)= \int_{\Omega}\mu \nu
\d x \d y \d z$ and $(u_{ijk},v_{ijk})=
\int_{\Omega}u_{ijk}v^{ijk}\d x \d y \d z$ for the spaces of
scalar fields and triple indexed fields on $\Omega$. The $L_{2}$
norms naturally associated with these scalar products are $\| \mu
\|^{2}=(\mu,\mu)$ and $\|u\|^{2}=(u_{ijk},u_{ijk})$. The energy of
system (\ref{eqA1}) is defined as
\begin{equation*}
\epsilon_{1} = \frac{1}{2}(\| \partial_{t} A \|^{2} + \| \partial_{l} A
\|^{2}).
\end{equation*}
Similar to Section~4, by differentiating $\epsilon_{1}$ in time,
integrating terms with spatial derivatives by parts, and using
(\ref{eqA1}), we obtain
\begin{equation}
\label{energy0}
\partial_{t}\epsilon_{1} =\int_{\partial\Omega} (\frac{\partial}{\partial n} A_{ij})(\partial_{t} A^{ij}) \d
\sigma,
\end{equation}
Since the right side of (\ref{energy0}) is zero for either
(\ref{ccbc1}) or (\ref{ccbc2}), we conclude that $\epsilon_{1}$,
and therefore, $A$ remains bounded.

Similarly, by differentiating (\ref{LBSSN3}) and substituting (\ref{LBSSN2})
for $\partial_{t}\alpha$ one derives the equation for $\kappa$
\begin{equation}
\label{eqkappa1}
\partial^{2}_{t}\kappa = \partial^{l}\partial_{l}\kappa.
\end{equation}
The boundary conditions (\ref{ccbc1}) or (\ref{ccbc2}) imply
trivial Neumann and Dirichlet data on $\kappa$ correspondingly.

Finally, assuming the shift perturbation $\beta$ is known, the
matrix $A$ is computed from (\ref{eqA1}), and $\kappa$ is
determined from (\ref{eqkappa1}), variables $\varphi$, $\alpha$,
$\tilde{\gamma}$, $\Gamma$ can be determined from (\ref{LBSSN1}),
(\ref{LBSSN2}), (\ref{LBSSN4}), (\ref{LBSSN6}) by integration in
time.

\section{Initial boundary value problem for the linearized
         BSSN. Prescribed and Gamma-freezing shift}

Let system (\ref{LBSSN1})--(\ref{LBSSN6}) be provided with
relevant initial data, and $\beta$ be given. If the boundary
conditions for $A$ and $\kappa$ are taken in either form
(\ref{ccbc1}) or (\ref{ccbc2}) (conditions (\ref{ccbc3}) and
(\ref{ccbc6}) can be formally imposed and treated similarly but
contain derivatives of the unknown fields), then boundary
conditions on variables $\tilde{\gamma}$, $\varphi$, $\alpha$, and
$\Gamma$ can be obtained by integration of (\ref{LBSSN4}),
(\ref{LBSSN1}), (\ref{LBSSN2}), (\ref{LBSSN6}) as (the projection
operator $N^{pq}_{ij}$ is defined below)
\begin{gather}
N^{pq}_{ij}\tilde{\gamma}_{pq} +
(\delta^{p}_{i}\delta^{q}_{j} - N^{pq}_{ij}) \frac{\partial}{\partial n}\tilde{\gamma}_{pq}
= N^{pq}_{ij}\tilde{\gamma}_{pq}(0)\hspace{10mm} \nonumber \\
{} + (\delta^{p}_{i}\delta^{q}_{j} - N^{pq}_{ij}) \frac{\partial}{\partial
n}\tilde{\gamma}_{pq}(0)+\int_{0}^{t}(N^{pq}_{ij} \nonumber \\
\label{bc5}
\quad {}+(\delta^{p}_{i}\delta^{q}_{j}-N^{pq}_{ij})\frac{\partial}{\partial n})
(2\partial_{(p}\beta_{q)}-\frac{2}{3}\delta_{pq} \partial^{s} \beta_{s}), \\
\mu \varphi+(1-\mu)\frac{\partial}{\partial n} \varphi
=\mu \varphi(0) + (1-\mu)\frac{\partial}{\partial n} \varphi(0)\hspace{10mm} \nonumber \\
\label{bc6}
\qquad {} +\int_{0}^{t} \frac{1}{6}(\mu-(1-\mu)\frac{\partial}{\partial
n})\partial^{l}\beta_{l},\\
\label{bc7}
\mu \alpha + (1-\mu)\frac{\partial}{\partial n} \alpha
=\mu \alpha(0) + (1-\mu)\frac{\partial}{\partial n} \alpha(0),\\
n^{i}\Gamma_{i}=n^{i}\Gamma_{i}(0) + \int_{0}^{t}
[-\frac{4}{3}\frac{\partial}{\partial n} \kappa + \frac{1}{3}\frac{\partial}{\partial n}\partial^{p}\beta_{p}
     +\partial^{p}\partial_{p} n^{i}\beta_{i}], \nonumber \\
\label{bc8}
\tau^{i}\Gamma_{i}=\tau^{i}\Gamma_{i}(0) + \int_{0}^{t}
[-\frac{4}{3}\frac{\partial}{\partial \tau} \kappa + \frac{1}{3}\frac{\partial}{\partial \tau}\partial^{p}\beta_{p}
     +\partial^{p}\partial_{p} \tau^{i}\beta_{i}].
\end{gather}
The projection operator $N^{pq}_{ij}$ corresponding to (\ref{ccbc1})
is
\begin{equation*}
N^{pq}_{ij} = 2n^{(p}m^{q)}n_{(i}m_{j)} + 2n^{(p}l^{q)}n_{(i}l_{j)}
\end{equation*}
and the one corresponding to (\ref{ccbc2}) is
\begin{gather*}
N^{pq}_{lk}=2l^{(p}m^{q)}l_{(i}m_{j)}+
(1/2)(l^{p}l^{q}-m^{p}m^{q})(l_{i}l_{j}-m_{i}m_{j})\\
+(1/6)(2n^{p}n^{q}-l^{p}l^{q}-m^{p}m^{q})(2n_{i}n_{j}-l_{i}l_{j}-m_{i}m_{j}),
\end{gather*}
$\mu=0$ corresponds to (\ref{ccbc1}) and $\mu=1$ to (\ref{ccbc2}). In
(\ref{bc8}), $\tau_{i}$ stands for vectors $l_{i}$, $m_{i}$.
Unless the Neumann data is given on $\kappa$, the equation
(\ref{bc8}) couples $\Gamma$ and $\kappa$ through
an integral equation. If the Neumann data is specified for
$\kappa$, for example in (\ref{ccbc1}), the last
equation can be replaced with
\begin{gather*}
\tau^{i}\frac{\partial}{\partial n}\Gamma_{i}=\tau^{i}\frac{\partial}{\partial n}\Gamma_{i}(0)
+ \int_{0}^{t} [-\frac{4}{3}\frac{\partial}{\partial \tau} \frac{\partial}{\partial
n}\kappa\\
+\frac{\partial}{\partial n}(\frac{1}{3}\frac{\partial}{\partial \tau}\partial^{p}\beta_{p}
     +\partial^{p}\partial_{p}\tau^{i}\beta_{i})].
\end{gather*}

\begin{thm}
Let $A$ and $\kappa$ are smooth solutions of (\ref{eqA1}) and
(\ref{eqkappa1}) corresponding to boundary data (\ref{ccbc1}) (or
(\ref{ccbc2})) and the initial data $A(0)$, $\kappa(0)$,
$\partial_{t} A(0)$, $\partial_{t} \kappa(0)$ (the last two are
determined from equations (\ref{LBSSN5}) and (\ref{LBSSN3}) and the initial
data $\tilde{\gamma}(0)$, $\Gamma(0)$, $\alpha(0)$, and $\varphi(0)$ as
\begin{gather}
\partial_{t}A_{ij}(0)
 = -\frac{1}{2}\partial^{l}\partial_{l} \tilde{\gamma}_{ij}(0)
     + \partial_{(i}\Gamma_{j)}(0) - 2\partial_{i}\partial_{j}\varphi(0) \nonumber \\
\label{eqA2}
{}-2\delta_{ij}\partial^{l}\partial_{l}\varphi(0)
      -\partial_{i}\partial_{j}\alpha(0)+\frac{1}{3}\delta_{ij}\partial^{l}\partial_{l}\alpha(0),\\
\label{eqkappa2}
\partial_{t}\kappa(0)= -\partial^{l}\partial_{l}\alpha(0). \quad
\mbox{)}
\end{gather}
Then a solution to
(\ref{LBSSN1})--(\ref{LBSSN6}) satisfying boundary data (\ref{ccbc1}) (or (\ref{ccbc2})),
(\ref{bc5})--(\ref{bc8}) is given by
\begin{gather}
\varphi
= \varphi(0)+\int_{0}^{t}
[-\frac{1}{6}\kappa+\frac{1}{6}\partial^{l}\beta_{l}],\quad
\alpha=\alpha(0)-\int_{0}^{t}\kappa, \nonumber\\
\tilde{\gamma}_{ij}=\tilde{\gamma}_{ij}(0)+
\int_{0}^{t}[-2A_{ij}+2\partial_{(i}\beta_{j)}
     -\frac{2}{3}\delta_{ij} \partial^{l} \beta_{l}],\nonumber \\
\label{eqXXX1}
\Gamma_{i}=\Gamma_{i}(0)+ \int_{0}^{t}[-\frac{4}{3}\partial_{i}\kappa +
\frac{1}{3}\partial_{i}\partial^{p}\beta_{p} +\partial^{p}\partial_{p}\beta_{i}].
\end{gather}
Moreover, if $A$ and $\kappa$ satisfy constraint equation
(\ref{LBSSN8}), then $\tilde{\gamma}$, $\Gamma$, and $\varphi$
defined by (\ref{eqXXX1}), satisfy constraints (\ref{LBSSN7}) and
(\ref{LBSSN9}) as long as they satisfy them at the initial time.
\end{thm}
\begin{proof}
Equations (\ref{LBSSN1}), (\ref{LBSSN2}), (\ref{LBSSN4}), and (\ref{LBSSN6}) are verified by
substitution. Replacing $\tilde{\gamma}$, $\Gamma$, $\varphi$, and $\alpha$ in (\ref{LBSSN5}) and (\ref{LBSSN3})
by their expressions from (\ref{eqXXX1}) and using (\ref{eqA2}) and (\ref{eqkappa2}) we
obtain
\begin{equation*}
\partial_{t}A_{ij}=\partial_{t}A_{ij}(0)
+\int_{0}^{t}\partial^{l}\partial_{l} A_{ij},\
\partial_{t}\kappa = \partial_{t}\kappa(0) +
\int_{0}^{t}\partial^{l}\partial_{l}\kappa
\end{equation*}
which is a consequence of (\ref{eqA1}) and (\ref{eqkappa1}).
Substituting (\ref{eqXXX1}) into (\ref{bc5})--(\ref{bc8}) we
verify the boundary conditions.

Now consider constraints (\ref{LBSSN7}). Replacing
$\tilde{\gamma}$, $\varphi$, and $\Gamma$ with their expressions
from (\ref{eqXXX1}) we obtain
\begin{gather}
\partial^{l}\partial^{j}\tilde{\gamma}_{lj}-8\partial^{l}\partial_{l}\varphi=
\partial^{l}\partial^{j}\tilde{\gamma}_{lj}(0)-8\partial^{l}\partial_{l}\varphi(0)
\nonumber \\
-\int_{0}^{t}2\partial^{j}(\partial^{l}A_{jl}-\frac{2}{3}\partial_{j}k),
\nonumber \\
\label{eqXXX2}
\partial^{l}\Gamma_{l}-8\partial^{l}\partial_{l}\varphi=
\partial^{l}\Gamma_{l}(0)-8\partial^{l}\partial_{l}\varphi(0),
\end{gather}
from which it follows that (\ref{LBSSN7}) is met as long as it is
satisfied initially and (\ref{LBSSN8}) is true. Constraint (\ref{LBSSN9})
follows similarly.
\end{proof}

The situation is similar when $\beta$ is to be determined from
the gamma-freezing condition $\partial_{t}\Gamma_{i}=0$ which
yields an elliptic equation for $\beta$ that can be solved at each
time step
\begin{equation}
\label{beta2}
\frac{1}{3}\partial_{i}\partial^{p}\beta_{p} +
\partial^{p}\partial_{p}\beta_{i}-\frac{4}{3}\frac{\partial}{\partial n}\kappa=0.
\end{equation}
The boundary conditions on $\beta$ can be taken, for example,
\begin{equation}
\label{bc9}
\beta_{i}m^{i}=0, \quad
\beta_{i}l^{i}=0, \quad
\frac{\partial}{\partial n}\beta_{i}n^{i}=0,
\end{equation}
or,
\begin{equation}
\label{bc10}
\frac{\partial}{\partial n}\beta_{i}m^{i}=0, \quad
\frac{\partial}{\partial n}\beta_{i}l^{i}=0, \quad
\beta_{i}n^{i}=0.
\end{equation}

It is beneficial for the computation purposes to replace the
elliptic equation (\ref{beta2}) with the hyperbolic equation  \cite{BTJ04}
\begin{equation}
\label{beta3}
\partial^{2}_{t}\beta=\frac{1}{3}\partial_{i}\partial^{p}\beta_{p} +
\partial^{p}\partial_{p}\beta_{i}-\frac{4}{3}\frac{\partial}{\partial
n}\kappa,
\end{equation}
which corresponds to a dynamic gamma-freezing condition
$\partial_{t}\Gamma_{i}=\partial^{2}_{t}\beta_{i}$. In
this case, in addition to (\ref{bc8}) and (\ref{bc9}), one can
consider either of two sets of the radiative boundary conditions
\begin{gather*}
\beta_{i}m^{i}=0, \quad
\beta_{i}l^{i}=0, \quad
(\partial_{t}\beta_{i}+\frac{\partial}{\partial n}\beta_{i})n^{i}=0,\nonumber \\
(\partial_{t}\beta_{i}+\frac{\partial}{\partial n}\beta_{i})m^{i}=0, \quad
(\partial_{t}\beta_{i}+\frac{\partial}{\partial n}\beta_{i})l^{i}=0, \quad
\beta_{i}n^{i}=0.
\end{gather*}

After the boundary conditions for $\beta$, $A$, and $\kappa$
are chosen, one can set $\Gamma_{i}-\Gamma(0)_{i}=0$ on the boundary, then
compatible conditions for the rest of the variables follow from
integration of the system as in the examples above.

\begin{center}
\textbf{Acknowledgment}
\end{center}

The author thanks D.~Arnold for helpful suggestions at the earlier
stages of this paper, O.~Sarbach and G.~Nagy for pointing out an
error in the first version of the paper, and L.~Lindblom and
M.~Scheel for fruitful discussions and interest in this work. A
special thanks is to H.~Pfeiffer for reading the final draft of the
manuscript and making useful comments.

\section{Appendix: Energy estimates for the BSSN system with boundaries}

We use differentiation in time to propose boundary conditions
(\ref{ccbc1}), (\ref{ccbc2}) (and the associated conditions
(\ref{bc5})--(\ref{bc8})) in Sections~4--6. Our derivation,
however, strongly relies on the linearization assumption and does
not extend to the nonlinear case.%
\footnote{The difficulty appears to be the extra
derivatives of the inverse metric resulting from differentiation
of (\ref{tildeA2}) that contaminate principal part and break the
similarity with the linear case.}
In other words, one could have proposed boundary conditions for
the nonlinear case on the base of (\ref{ccbc1}) (or
(\ref{ccbc2})), but one will not be able to prove well-posedness
of the new conditions by repeating the argument of Sections~4--6. In
this appendix we try to fix this flaw by establishing a well posed
energy estimate without reduction to second order in time. We use
approach proposed by Gundlach and Martin-Garcia
\cite{GMG04a,GMG04b} and modify the proof for the case of a
bounded domain.

Following \cite{GMG04b}, we use the densitized lapse
\begin{equation}
\label{a3}
a=\e^{6\varphi}Q,
\end{equation}
which results in significant simplifications in equations. The
result, however, is expected to  extend to harmonic slicing and
$k$-driving slicing as well.

We illustrate the idea in the linear case, first. Under the
densitized lapse assumption, equation (\ref{LBSSN2}) is replaced
with the linearized densitized lapse condition, the latter in the
special case $Q\equiv 1$, reduces to $\alpha=6\varphi$. Assuming
for further simplicity zero shift perturbation ($\beta_{i}=0$),
the equations (\ref{LBSSN1}), (\ref{LBSSN3})--(\ref{LBSSN6}), are
restated as
\begin{eqnarray}
\label{RBSSN1}
\partial_{t}\varphi
 &=& -\frac{1}{6}\kappa, \\
\label{RBSSN2}
\partial_{t}\kappa
 &=&-6\partial^{l}\partial_{l}\varphi, \\
\label{RBSSN3}
\partial_{t}\tilde{\gamma}_{ij}
 &=& -2A_{ij},\\
\label{RBSSN4}
\partial_{t}A_{ij}
 &=& -\frac{1}{2}\partial^{l}\partial_{l} \tilde{\gamma}_{ij}
     + \partial_{(i}\Gamma_{j)} - 8\partial_{i}\partial_{j}\varphi,\\
\label{RBSSN5}
\partial_{t}\Gamma_{i}
 &=& -\frac{4}{3}\partial_{i}\kappa.
\end{eqnarray}
Taking scalar product of (\ref{RBSSN2}) with $\kappa$, integrating
the result by parts, and using (\ref{RBSSN1}) to replace $\kappa$
with $\partial_{t}\varphi$, we obtain
\begin{equation}
\label{eest1}
\frac{1}{2}\partial_{t}[\| \kappa \|^{2}+36\|\partial_{l} \varphi
\|^{2}]= -6 \int_{\partial \Omega} (\frac{\partial}{\partial
n}\varphi)\kappa.
\end{equation}
Also, we conclude from (\ref{RBSSN1}) that $36\| \partial_{t}\varphi \|^{2}=\| \kappa
\|^{2}$.

From (\ref{RBSSN1}) and (\ref{RBSSN5}) we observe that
$\partial_{t}(\Gamma_{i}-8\partial_{i}\varphi)=0$, which implies
that
\begin{equation}
\label{eest2}
\partial_{t}\|\Gamma_{i}-8\partial_{i}\varphi\|^2=0,\quad
(\partial_{t}(\Gamma_{i}-8\partial_{i}\varphi),\partial^{l}\tilde{\gamma}_{li})=0.
\end{equation}

Next we rewrite the right side of (\ref{RBSSN4}) in divergence form,
\begin{equation*}
\partial_{t}A_{ij} = \partial^{l}[-\frac{1}{2}\partial_{l} \tilde{\gamma}_{ij}
     +\delta_{l(i}(\Gamma_{j)} - 8\partial_{j)}\varphi)],
\end{equation*}
then, take scalar product with $A_{ij}$, and integrate by parts:
\begin{gather*}
\frac{1}{2}\partial_{t}\|A\|^{2}+
( -\frac{1}{2}\partial_{l} \tilde{\gamma}_{ij}
  + \delta_{l(i}(\Gamma_{j)} - 8\partial_{j)}\varphi), \partial_{l}
  A_{ij})\nonumber \\
=\int_{\partial \Omega} [-\frac{1}{2}\frac{\partial}{\partial n} \tilde{\gamma}_{ij}
     + n_{(i}(\Gamma_{j)} - 8\partial_{j)}\varphi)] A^{ij}.
\end{gather*}
Replacing $A_{ij}$ with (\ref{RBSSN3}) in the second term and
using $\partial_{t}[\delta_{l(i}(\Gamma_{j)} - 8\partial_{j)}\varphi)]=0$,
we replace the last identity with
\begin{gather}
\frac{1}{2}\partial_{t} [ \|A\|^{2}+ \| \frac{1}{2}\partial_{l} \tilde{\gamma}_{ij}
  - \delta_{l(i}(\Gamma_{j)} - 8\partial_{j)}\varphi)\|^{2} ] \nonumber \\
\label{eest4}
=\int_{\partial \Omega} [-\frac{1}{2}\frac{\partial}{\partial n} \tilde{\gamma}_{ij}
     + n_{(i}(\Gamma_{j)} - 8\partial_{j)}\varphi)] A^{ij}.
\end{gather}

From (\ref{eest1})--(\ref{eest4}) one observes that the energy
\begin{gather}
\epsilon=\|\kappa\|^{2}+36\| \partial_{l}\varphi \|^{2}+
\|\Gamma_{l}-8\partial_{l}\varphi\|^{2}+\| A\|^{2}\nonumber \\
\label{energy1}
+\|\frac{1}{2}\partial_{l}
\tilde{\gamma}_{ji}-\delta_{l(i}(\Gamma_{j)} - 8\partial_{j)}\varphi)\|^{2},
\end{gather}
has its growth determined by the boundary terms:
\begin{gather}
\partial_{t}\epsilon
= -6\int_{\partial \Omega} (\frac{\partial}{\partial n}\varphi)\kappa
  -\int_{\partial \Omega}(\frac{\partial}{\partial n}\tilde{\gamma}_{ij})A^{ij} \nonumber \\
\label{eest5}
  +2\int_{\partial \Omega} n_{(i}(\Gamma_{j)} - 8\partial_{j)}\varphi)A^{ij}.
\end{gather}

Expression (\ref{eest5}) can be used to propose examples of new
energy (or constraint) preserving boundary conditions for the
linearized BSSN formulation. However, here we will use
(\ref{eest5}) to discuss the meaning of conditions (\ref{ccbc1})
and (\ref{ccbc2}) proposed in Section~6. Under assumption of
(\ref{ccbc1}) ((\ref{ccbc2}) can be treated similar) and the
associated conditions (\ref{bc5}), (\ref{bc6}), (\ref{bc8}),
expression (\ref{eest5}) reduces to
\begin{gather}
\partial_{t}\epsilon
= \int_{\partial \Omega}[-6\frac{\partial}{\partial n}\varphi(0)\kappa
-\frac{1}{2} \frac{\partial}{\partial n}\tilde{\gamma}3(0)A3
-2\frac{\partial}{\partial n}\tilde{\gamma}4(0)A4\nonumber \\
\label{eest6}
- 6\frac{\partial}{\partial n}\tilde{\gamma}5(0)A5
+ 4(n^{i}\Gamma_{i}(0) - 8\frac{\partial}{\partial n}
\varphi(0))A5],
\end{gather}
where $\tilde{\gamma}$1--5 are the coefficients of decomposition
\begin{eqnarray*}
\label{tildegammaX}
\tilde{\gamma}_{ij}
&=& \tilde{\gamma}1\, (n_{(i}m_{j)})+ \tilde{\gamma}2\, (n_{(i}l_{j)}) + \tilde{\gamma}3\, (l_{(i}m_{j)}) \nonumber \\
& &{}+\tilde{\gamma}4\, (l_{i}l_{j}-m_{i}m_{j}) +
\tilde{\gamma}5\,(2n_{i}n_{j}-l_{i}l_{j}-m_{i}m_{j}).
\end{eqnarray*}
The energy (\ref{energy1}) is conserved if the initial data is
chosen as to satisfy $(\partial/\partial n) \varphi(0)=0$,
$\tilde{\gamma}12(0)=0$, and $(\partial/\partial n)
\tilde{\gamma}345(0)=0$ at the boundary (notice that $\Gamma_{i}(0)$ can not be
given freely but is expected to be subject to constraint
(\ref{LBSSN9})).

Similarly, conditions (\ref{ccbc2}) are constraint and energy
preserving for (\ref{RBSSN1})--(\ref{RBSSN5}) if
$\varphi=0$, $(\partial/\partial n) \tilde{\gamma}12(0)=0$, and
$\tilde{\gamma}345(0)=0$ on $\partial\Omega$.

We are now ready to establish the nonlinear analog of
(\ref{eest5}). Substituting (\ref{a3}) into the BSSN equations,
distributing and expanding covariant derivatives, we rewrite the
system,
\begin{align}
\label{DBSSN1}
\e^{4\varphi} \partial_{0}k
 &=-6a \partial^{p}\partial_{q}\varphi + F, \\
\label{DBSSN2}
\partial_{0} \varphi
 &= -\frac{1}{6}ak + \frac{1}{6}\partial_{l}\tilde{b}^{l},\\
\label{DBSSN3}
\partial_{0} \tilde{h}_{ij}
 &= -2a\tilde{A}_{ij}
     +2\tilde{h}_{l(i}\partial_{j)}\tilde{b}^{l}
     -\frac{2}{3} \tilde{h}_{ij}\partial_{l}\tilde{b}^{l},\\
\label{DBSSN4}
\e^{4\varphi}\partial_{0}\tilde{A}_{ij}
 &=-a\partial^{p}[\frac{1}{2}\partial_{p}\tilde{h}_{ij}
     -\tilde{h}_{p(i}(\tilde{\Gamma}_{j)}-8\partial_{j)}\varphi)]
     + G_{ij},\\
\label{DBSSN5}
\partial_{0}\tilde{\Gamma}_{i}
 &=-\frac{4}{3}a\partial_{i} k + S_{i}.
\end{align}
where
\begin{gather*}
F=6a\tilde{h}^{pq}\tilde{\Gamma}^{l}_{pq}\partial_{l}\varphi
-48a(\partial^{p}\varphi)(\partial_{p}\varphi)
-14\e^{6\varphi}(\partial^{p}\varphi)(\partial_{p}Q)\\
-\e^{6\varphi}(\tilde{D}^{p}\tilde{D}_{p}Q)
+\frac{1}{3}\e^{4\varphi}ak^{2}+\e^{4\varphi}a\tilde{A}^{pq}\tilde{A}_{pq},\\
G_{ij}=-a(\partial^{p}\tilde{h}_{p(i})(\tilde{\Gamma}_{j)}-8\partial_{j)}\varphi)
+8a\tilde{\Gamma}_{ij}^{p}\partial_{p}\varphi\\
-12a(\partial_{i}\varphi)(\partial_{j}\varphi)
+4a\tilde{h}_{ij}(\partial^{p}\varphi)(\partial_{p}\varphi)
-8\e^{6\varphi}(\partial_{(i}\varphi)(\partial_{j)}Q)\\
-\e^{6\varphi}(\tilde{D}_{i}\tilde{D}_{j}Q)
+\frac{8}{3}\e^{6\varphi}\tilde{h}_{ij}(\partial^{p}\varphi)(\partial_{p}Q)\\
+\frac{1}{3}\e^{6\varphi}\tilde{h}_{ij}(\tilde{D}^{p}\tilde{D}_{p}Q)
+\e^{4\varphi}W_{ij}.
\end{gather*}
Taking scalar product of (\ref{DBSSN1}) with $k$ and integrating
by parts in the right side, we get ($\tilde{n}_{i}$ is the outward
normal vector to the boundary in the sense of the conformal metric $\tilde{h}$),
\begin{gather*}
\int_{\Omega}(\partial_{0}k)k\e^{4\varphi}
=6\int_{\Omega} (\partial_{p}\varphi) (\partial_{q}(ak))h^{pq}
-6\int_{\partial\Omega} (\tilde{n}^{p}\partial_{p}\varphi)ak \nonumber \\
+\int_{\Omega} [6(\partial_{p}\varphi) ak \partial_{q}h^{pq}+Fk].
\end{gather*}
Substituting (\ref{DBSSN2}) for $ak$ in the second term and
re-grouping we obtain our first energy identity:
\begin{gather}
\label{nenerg1}
\frac{1}{2}\partial_{0}[\|k \|_{\ast}^{2}+36\|\partial_{l}\varphi\|^{2}]
=-6\int_{\partial \Omega}
(\tilde{n}^{p}\partial_{p}\varphi)ak+\int_{\Omega} H,
\end{gather}
where $\|k \|^2_{\ast}=\int_{\Omega} k^2\e^{4\varphi}$, and
\begin{gather*}
H=6(\partial_{p}\varphi)[\partial_{q}\partial_{s}\tilde{b}^{s}
  +6(\partial_{q}\tilde{b}^{s})(\partial_{s}\varphi)]\tilde{h}^{pq}\\
  +36(\partial_{p}\varphi)(\partial_{q}\varphi)(a\tilde{A}^{pq}-\partial^{(p}\tilde{b}^{q)}
  +\frac{1}{3}\tilde{h}^{pq}\partial_{s}\tilde{b}^{s})\\
  +k^{2}\e^{4\varphi}(-\frac{1}{3} ak +\frac{1}{3}\partial_{s}\tilde{b}^{s})
  +6(\partial_{p}\varphi) ak \partial_{q}\tilde{h}^{pq}+Fk.
\end{gather*}
Also, from (\ref{DBSSN2}) it follows that
\begin{equation}
\label{nenerg2}
\frac{1}{2}\partial_{0}\| \varphi \|^{2}
= \int_{\Omega}[-\frac{1}{6}ak
+\frac{1}{6}\partial_{l}\tilde{b}^{l}]\varphi.
\end{equation}
Next, we notice from (\ref{DBSSN1}) and (\ref{DBSSN5}) that
\begin{gather*}
\partial_{0}(\Gamma_{i}-8\partial_{i}\varphi) =
8a(\partial_{i}\varphi)k + \frac{4}{3}\e^{6\varphi}(\partial_{i}Q)k
-\frac{4}{3}\partial_{i}\partial_{s}\tilde{b}^{s}\\
-8(\partial_{i}\tilde{b}^{s})(\partial_{s}\varphi) + S_{i}.
\end{gather*}
Therefore,
\begin{equation}
\label{nenerg3}
\frac{1}{2} \partial_{0} \| \Gamma_{i}-8\partial_{i}\varphi \|^2 = \int_{\Omega} J,
\end{equation}
\begin{gather*}
J = [8a(\partial_{p}\varphi)k + \frac{4}{3}\e^{6\varphi}(\partial_{p}Q)k
-\frac{4}{3}\partial_{p}\partial_{s}\tilde{b}^{s}\\
-8(\partial_{p}\tilde{b}^{s})(\partial_{s}\varphi) + S_{p}]
(\tilde{\Gamma}_{q}-8\partial_{q}\varphi)\tilde{h}^{pq}\\
+\frac{1}{2}(\tilde{\Gamma}_{p}-8\partial_{p}\varphi)(\tilde{\Gamma}_{q}-8\partial_{q}\varphi)
 [2a\tilde{A}^{pq}-\partial^{(p}\tilde{b}^{q)}+\frac{2}{3}\tilde{h}^{pq}\partial_{s}\tilde{b}^{s}].
\end{gather*}
Finally, taking scalar product of (\ref{DBSSN4}) with $A$,
integrating by parts in the right side, using (\ref{DBSSN3}) to replace $A$
with $\partial_{0}h$, and re-grouping, we
derive our last energy identity:
\begin{gather}
\frac{1}{2}\partial_{0} [\|\tilde{A}\|^2_{\ast}+
\|\frac{1}{2}\partial_{l}\tilde{h}_{ij}-\tilde{h}_{l(i}(\tilde{\Gamma}_{j)}-8\partial_{j)}\varphi)\|^{2}]\nonumber \\
\label{nenerg4}
= -\int_{\partial \Omega} [\frac{1}{2} (\tilde{n}^{p}\partial_{p}\tilde{h}_{ij})
  -\tilde{n}_{(i}(\tilde{\Gamma}_{j)}-8\partial_{j)}\varphi)]a\tilde{A}^{ij}+\int_{\Omega}K,
\end{gather}
where $\|A\|^{2}_{\ast}=\int_{\Omega}A_{ij}A^{ij}\e^{4\varphi}$, and
\begin{gather*}
K=[\frac{1}{2}\partial_{p}\tilde{h}_{ij}-\tilde{h}_{p(i}(\tilde{\Gamma}_{j)}-8\partial_{j)}\varphi)]\\
\times \{[\partial_{q}(\tilde{h}_{s(m}\partial_{n)}\tilde{b}^{s}
-\frac{1}{3}\tilde{h}_{mn}\partial_{s}\tilde{b}^{s})+(\partial_{q}\tilde{b}^{s})(\partial_{s}\tilde{h}_{mn}) \\
-(-2a\tilde{A}_{q(m}+\tilde{h}_{lq}\partial_{(m}\tilde{b}^{l}+\tilde{h}_{s(m}\partial_{|q|}\tilde{b}^{s}
-\frac{2}{3}\tilde{h}_{q(m}\partial_{|s|}\tilde{b}^{s})
(\tilde{\Gamma}_{n)}\\
-8\partial_{n)}\varphi)
-\tilde{h}_{q(m}(8a(\partial_{n)}\varphi)k + \frac{4}{3}\e^{6\varphi}(\partial_{n)}Q)k
-\frac{4}{3}\partial_{n)}\partial_{s}\tilde{b}^{s} \\
-8(\partial_{n)}\tilde{b}^{s})(\partial_{s}\varphi) +
S_{n)})]\tilde{h}^{pq}\tilde{h}^{im}\tilde{h}^{jn}\\
+a\tilde{A}_{mn}\partial_{q}(\tilde{h}^{pq}\tilde{h}^{im}\tilde{h}^{jn})\}+G_{ij}A^{ij}.
\end{gather*}
Notice that boundary terms in (\ref{nenerg1}) and (\ref{nenerg4})
differ from that of linear case in spatial metric only. Notice
also, that right sides of (\ref{nenerg1})--(\ref{nenerg4}) are
combinations of $\varphi$, $k$, $\tilde{h}$, $\tilde{A}$,
$\tilde{\Gamma}$, $\partial \varphi$, $\partial \tilde{h}$, but
not derivatives of $k$, $\tilde{A}$, and $\tilde{\Gamma}$, which
implies that (\ref{nenerg1})--(\ref{nenerg4}) is a closed estimate
and may be proposed for proving local well-posedness of
initial-boundary value problem for the BSSN system. In this
derivation, we assumed that covariant components of the conformal
shift vector, $\tilde{b}^{s}$, are prescribed, otherwise, terms
$\partial_{i}\partial_{j}\tilde{b}^{s}$ has to be expanded to
ensure that they do not contribute to the principal part of the
equations.

\bibliographystyle{apsrev}
\bibliography{n032405}
\end{document}